\documentclass[twocolumn, showkeys, showpacs, nofootinbib, amsfonts, amssymb, amsmath]{revtex4-2} 

\usepackage[english]{babel}
\usepackage[utf8]{inputenc}
\usepackage[colorinlistoftodos, color=green!40, prependcaption]{todonotes}
\usepackage{bm}
\usepackage{float} 
\usepackage{amsthm}
\usepackage{mathtools}
\usepackage{physics}
\usepackage{xcolor}
\usepackage{graphicx}
\usepackage[left=23mm,right=13mm,top=35mm,columnsep=15pt]{geometry} 
\usepackage{adjustbox}
\usepackage{placeins}
\usepackage[T1]{fontenc}
\usepackage{lipsum}
\usepackage{csquotes}
\usepackage[UTF8,scheme=plain,fontset=windows]{ctex}
\usepackage[pdftex, pdftitle={Article}, pdfauthor={Author}]{hyperref} 
\hypersetup{
    colorlinks=true,
    linkcolor=blue,
    citecolor=blue,
    urlcolor=blue
}

\begin{document}
\title{Swarming Lattice in Frustrated Vicsek-Kuramoto Systems}

\author{
    Yichen~Lu$^{1,2,\dagger}$, Yingshan Guo$^{1,4,\dagger}$, Yiyi Zhang$^{1,3}$, Tong Zhu$^{1,3}$, and Zhigang~Zheng$^{1,3,*}$
}
\footnotetext{$^{\dagger}$These authors contributed equally to this work.}
\footnotetext{$^*$Corresponding author. zgzheng@hqu.edu.cn}
\affiliation{ 
    $^1$Institute of Systems Science, Huaqiao University, Xiamen 361021, China\\
    $^2$School of Mathematical Sciences, Huaqiao University, Quanzhou, 362021, China\\
    $^3$College of Information Science and Technology, Huaqiao University, Xiamen 361021, China\\
    $^4$College of Mechanical Engineering and Automation Science and Technology, Huaqiao University, Xiamen 361021, China
}	

\date{\today}

\begin{abstract}
    We introduce a frustration parameter $\alpha$ into the Vicsek-Kuramoto systems of self-propelled particles. While the system exhibits conventional synchronized states, such as global phase synchronization and swarming, for low frustration ($\alpha < \pi/2$), beyond the critical point $\alpha = \pi/2$, a Hopf–Turing bifurcation drives a transition to a resting hexagonal lattice, accompanied by spatiotemporal patterns such as vortex lattices and dual-cluster lattices with oscillatory unit-cell motions. Lattice dominance is governed by coupling strength and interaction radius, with a clear parametric boundary balancing pattern periodicity and particle dynamics. Our results demonstrate that purely orientational interactions are sufficient to form symmetric lattices, challenging the necessity of spatial forces and illuminating the mechanisms driving lattice formation in active matter systems.
\end{abstract}


\maketitle

The collective motion of self-propelled particles occurs in systems from bird flocks \cite{PhysRevLett.75.1226,doi:10.1073/pnas.1118633109,Qiu2015} and fish schools \cite{doi:10.1073/pnas.0711437105,Becker2015} to synthetic active matter like Janus particles \cite{B718131K,doi:10.1021/cr300089t,Campbell2019,doi:10.1021/acs.accounts.8b00243}. These exhibit behaviors such as clustering \cite{Zhou_2023,PhysRevLett.110.238301,Ishimoto2018,C5SM01061F}, lane formation \cite{PhysRevE.94.052603,Kogler_2015,C0SM01343A}, swirling \cite{PhysRevLett.100.058001,riedel2005self,PhysRevLett.93.098103,Sumino2012,PhysRevLett.114.168001,PhysRevLett.127.238001,PhysRevLett.119.058002,LU2025115794}, turbulence \cite{doi:10.1073/pnas.1710188114,PhysRevE.97.022613,PhysRevE.98.022603}, and crystallization or lattice structures \cite{doi:10.1126/sciadv.aat7779,PhysRevE.97.052615,PhysRevLett.108.268303,doi:10.1128/AEM.68.12.6310-6320.2002,10.1016/j.femsle.2005.03.036,doi:10.1126/science.1230020,PhysRevE.109.024602,Xu2024,1605401,PhysRevE.111.045423,9998071,Sumino2012,PhysRevLett.114.168001}. Many are captured by Vicsek-type models, where particles align with neighbors under noise \cite{PhysRevLett.75.1226}, but these usually rely on spatial interactions like repulsion, attraction \cite{1605401,PhysRevE.111.045423,9998071} or stress-induced fluidization \cite{Xu2024} common in high-density systems with strong steric effects. 
Purely orientational interactions seldom form symmetric spatial structures alone, raising the question of whether lattice structures can emerge in systems where collective scales exceed individual scales (e.g., bird flocks) or in dilute regimes (e.g., gas molecules)?

Since the 2010s, the Vicsek model has been linked to the Kuramoto model \cite{RevModPhys.77.137,RODRIGUES20161,2019Emergence} by equating motion direction with oscillator phase \cite{PhysRevLett.108.248101,doi:10.1142/S0218202513400095}, forming a Vicsek-Kuramoto framework. Recent systems combining self-propulsion and phase interactions reveal chirality \cite{PhysRevLett.119.058002,PhysRevLett.127.238001,LU2025115794,Lu_2025}, memory effects \cite{Sumino2012,PhysRevLett.114.168001}, chimeras \cite{PhysRevE.98.032219,PhysRevE.102.022604}, and nematic \cite{Sumino2012,PhysRevLett.114.168001,PhysRevLett.127.238001} or anti-aligning interactions \cite{PhysRevE.109.024602}, often self-organizing into complex patterns.
Evidence suggests purely orientational interactions can form symmetric structures; for example, microtubules form vortex lattices via nematic alignment \cite{Sumino2012}, modeled with angular memory \cite{PhysRevLett.114.168001}, and anti-aligning interactions can yield traveling hexagonal lattices \cite{PhysRevE.109.024602}. 
However, these often require specific interactions or effects.

In this work, we investigate a minimal model of self-propelled particles with frustrated local alignment, inspired by the Kuramoto–Sakaguchi model for frustrated oscillators \cite{10.1143/PTP.76.576}. The frustration, which can be interpreted as a phase lag arising from internal or external factors, is commonly observed in real-world systems such as Josephson synchronization \cite{PhysRevLett.76.404}, optical \cite{Larger2015}, electrical \cite{PhysRevE.92.052912} and power grid \cite{Filatrella2008} systems.
By introducing a frustration parameter, we uncover a rich array of collective states, including synchronization, swarming, vortices, dual clusters, and dual-lane patterns. Remarkably, beyond a critical frustration threshold, the system transitions into a resting hexagonal lattice state and exhibits dominance for certain parameter ranges. We systematically map the phase diagram as a function of frustration, coupling strength, and interaction radius. Our analytical results illuminate the mechanisms driving lattice formation in active matter systems.

{\it Model and Order Parameters.}
Particles are described by position $\mathbf{r}_i=\left( x_i, y_i \right)$ and phase angle $\theta_i$, evolving as:
\begin{subequations} 
    \label{eq:totalDynamicsMeanField}
    \begin{align}
        \dot{\mathbf{r}}_i&=v\mathbf{p}\left( \theta _i \right)+\sqrt{2D}\bm{\xi}_i\;\label{eq:dotR},
        \\
        \dot{\theta}_i&=\omega_i+\frac{K}{\left| A_i \right|}\sum_{j\in A_i}{F\left( \theta _j-\theta _i \right) }+\sqrt{2D_r}\eta _i\;\label{eq:dotTheta},
    \end{align}
\end{subequations}
for $i=1,2,\ldots,N$, where $N$ is the population size, $v$ is self-propulsion speed, $\mathbf{p}\left( \theta _i \right) =\left( \cos \theta _i,\sin \theta _i \right)$ is the unit vector in the direction $\theta_i$, $\omega_i$ is the intrinsic frequency, $A_i\left( t \right) =\left\{ j: \left| \mathbf{r}_i\left( t \right) -\mathbf{r}_j\left( t \right) \right|\leqslant d_0 \right\}$ is the set of neighbors within coupling radius $d_0$ and $|A_i|$ its cardinality, $K \left(\geqslant 0\right)$ is coupling strength, $\bm{\xi}$ and $\eta$ are Gaussian white noises with zero mean and unit variance, and $D$ and $D_r$ are translational and rotational diffusion coefficients, respectively. 
Here, the interaction function is defined as $F\left( \theta \right) =\sin \left( \theta +\alpha \right) -\sin \alpha $, where $\alpha$ is frustration.
The term $-\sin\alpha$ ensures synchronization ($\theta_j = \theta_i$) remains an equilibrium \cite{10.1143/PTP.79.1069}. 
This system generalizes alignments \cite{PhysRevLett.119.058002,PhysRevResearch.1.023026,Escaff2020,PhysRevLett.127.238001,PhysRevLett.133.258302}, anti-alignments \cite{PhysRevE.109.024602,PhysRevE.110.024603}, and self-propelled chimera \cite{PhysRevE.98.032219,PhysRevE.102.022604}. It reduces to normal Vicsek–Kuramoto for $\alpha=0$ and anti-aligning interactions for $\alpha=\pi$.
We set $\omega_i=0$ (identical particles) for simplifying the theoretical analysis and $D=D_r=0$ (deterministic dynamics) for clarity; simulations with heterogeneous frequencies and noise are provided in the Supplemental Material \cite{mycomment2023}, which manifest qualitatively similar behaviors.

To quantify coordination, we define the single-particle distribution $\rho \left( \mathbf{r},\theta ,t \right) $, normalized as $\int_{L^2}{\mathrm{d}^2\mathbf{r}\int_0^{2\pi}{\rho\mathrm{d}\theta}}=1$, and derive marginalized densities:
\begin{subequations}
    \begin{align}
        \varrho \left( \mathbf{r},t \right)& =\int_0^{2\pi}{\rho \left( \mathbf{r},\theta ,t \right) \mathrm{d}\theta}\;,\\
        p\left( \theta ,t \right) & =\int_{L\times L}{\rho \left( \mathbf{r},\theta ,t \right) \mathrm{d}\mathbf{r}}\;,
    \end{align}
\end{subequations}
where $L$ is system size.
The homogeneous solution is $( \rho ,\varrho ,p ) =( \rho _0,\varrho _0,p_0 ) =\left( 1/( 2\pi L^2 ) ,1/L^2,1/(2\pi) \right) $. Standardized order parameters measure deviations:
\begin{equation}
    \rho _{\mathrm{std}}(t)=\frac{1}{1-\rho _0}\left[ \max_{\mathbf{r},\theta} \rho \left( \mathbf{r},\theta ,t \right) -\rho _0 \right]\;,
\end{equation}
with analogous $\varrho _{\mathrm{std}}$ and $p_{\mathrm{std}}$ for spatial condensation and phase polarization, respectively.
These range in $[0,1]$, with $0$ indicating uniformity and $1$ full condensation/polarization.

\begin{figure}
        \centering
        \includegraphics[width=7.5cm]{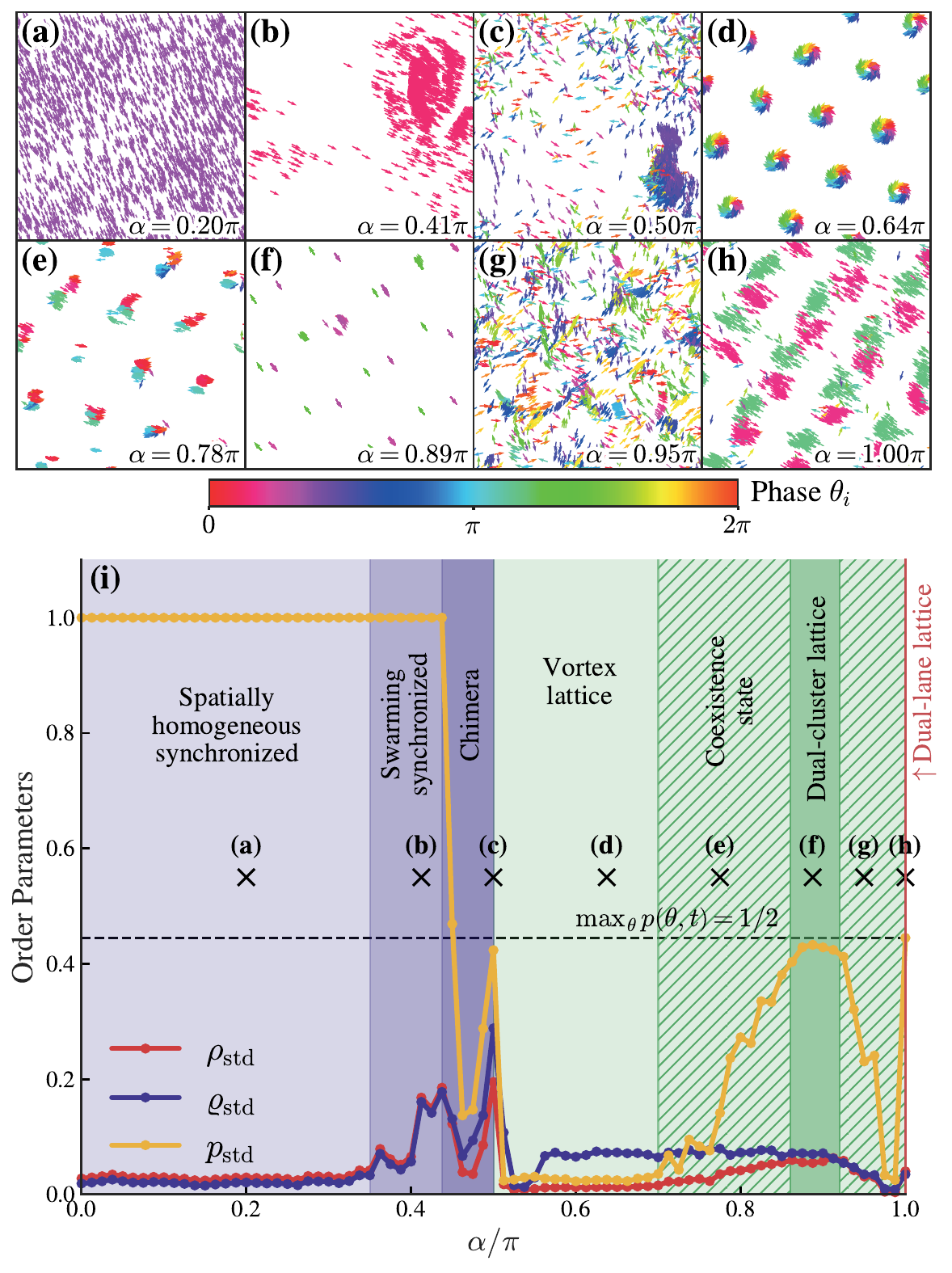}
        \caption{
            \label{fig:snapshotsAndPhaseDiagram}
            (a)-(h) Spatially homogeneous sync (a), swarming sync (b), chimera (c), vortex lattice (d), dual-cluster lattice (f), dual-lane lattice (h) states, and their coexistence states (e, g), at different frustration $\alpha$. Arrow orientation/color indicates instantaneous phase $\theta_i$.
            (i) Phase diagram and order parameters versus $\alpha$. Blue/green regions (divided by $\alpha=\pi/2$) indicate sync and lattice states, respectively. Dashed horizontal line shows $p_{\mathrm{std}}$ when $\max p(\theta, t)=1/2$ (dual-cluster sync). After reaching a steady state, order parameters were computed by averaging over 500 simulation steps.
            Other parameters: $K=20$, $d_0=1.55$, $L=7, N=2000, v=3$. 
        }
\end{figure}

{\it Lattice Structures and Spatiotemporal Patterns.} 
We now simulate the collective dynamics of $N$ frustrated self-propelled particles in a square domain of size $L\times L$ with periodic boundary conditions; using a time step of $\Delta t=0.005$.
At small $\alpha$, the system achieves global phase synchronization (Fig.~\ref{fig:snapshotsAndPhaseDiagram}(a)), with $p_{\mathrm{std}}=1$ and $\rho,\varrho_{\mathrm{std}}=0$, indicating aligned motion and spatial homogeneity. Increasing $\alpha$ leads to a swarming synchronized state (Fig.~\ref{fig:snapshotsAndPhaseDiagram}(b)), where particles form a coherent, moving cluster; $p_{\mathrm{std}}=1$ confirms polarization, while non-zero $\rho_{\mathrm{std}}$, $\varrho_{\mathrm{std}}$ reflect clustering. For larger $\alpha < \pi/2$, a chimera state emerges (Fig.~\ref{fig:snapshotsAndPhaseDiagram}(c)), mixing polarly ordered density cluster and incoherent low-density gas, as the phenomenology studied in \cite{PhysRevE.98.032219,PhysRevE.102.022604}.

For $\alpha > \pi/2$, lattice states emerge (Fig.~\ref{fig:snapshotsAndPhaseDiagram}(d)–(h)). Particles arrange into a hexagonal lattice, with $\rho_{\mathrm{std}}$ nearly constant for $0.55\pi<\alpha<0.9\pi$, indicating stable patterning in spatial density. 
Meanwhile, $p_{\mathrm{std}}$ decreases for $0.5\pi<\alpha<0.7\pi$, signifying a vortex lattice (Fig.~\ref{fig:snapshotsAndPhaseDiagram}(e)), similar to Refs \cite{Sumino2012,PhysRevLett.114.168001,Xu2024}. 
For $0.7\pi<\alpha<\pi$, $p(\theta, t)$ rises, marking a dual-cluster lattice (Fig.~\ref{fig:snapshotsAndPhaseDiagram}(f)) with two phase groups differing by $\pi$, leading to a bimodal distribution in $p(\theta, t)$ with $\max_{\theta} p(\theta, t)=1/2$.
Approaching $\alpha = \pi$, order parameters decrease, indicating disorder (Fig.~\ref{fig:snapshotsAndPhaseDiagram}(g)). At $\alpha=\pi$, anti-synchronization occurs (Fig.~\ref{fig:snapshotsAndPhaseDiagram}(h)), forming opposing particle lanes with fixed $\pi$ phase difference, as seen in the Vicsek model of anti-aligning interaction \cite{bisht2024laneslatticestructuresrepulsive}.

\begin{figure}
        \centering
        \includegraphics[width=7.5cm]{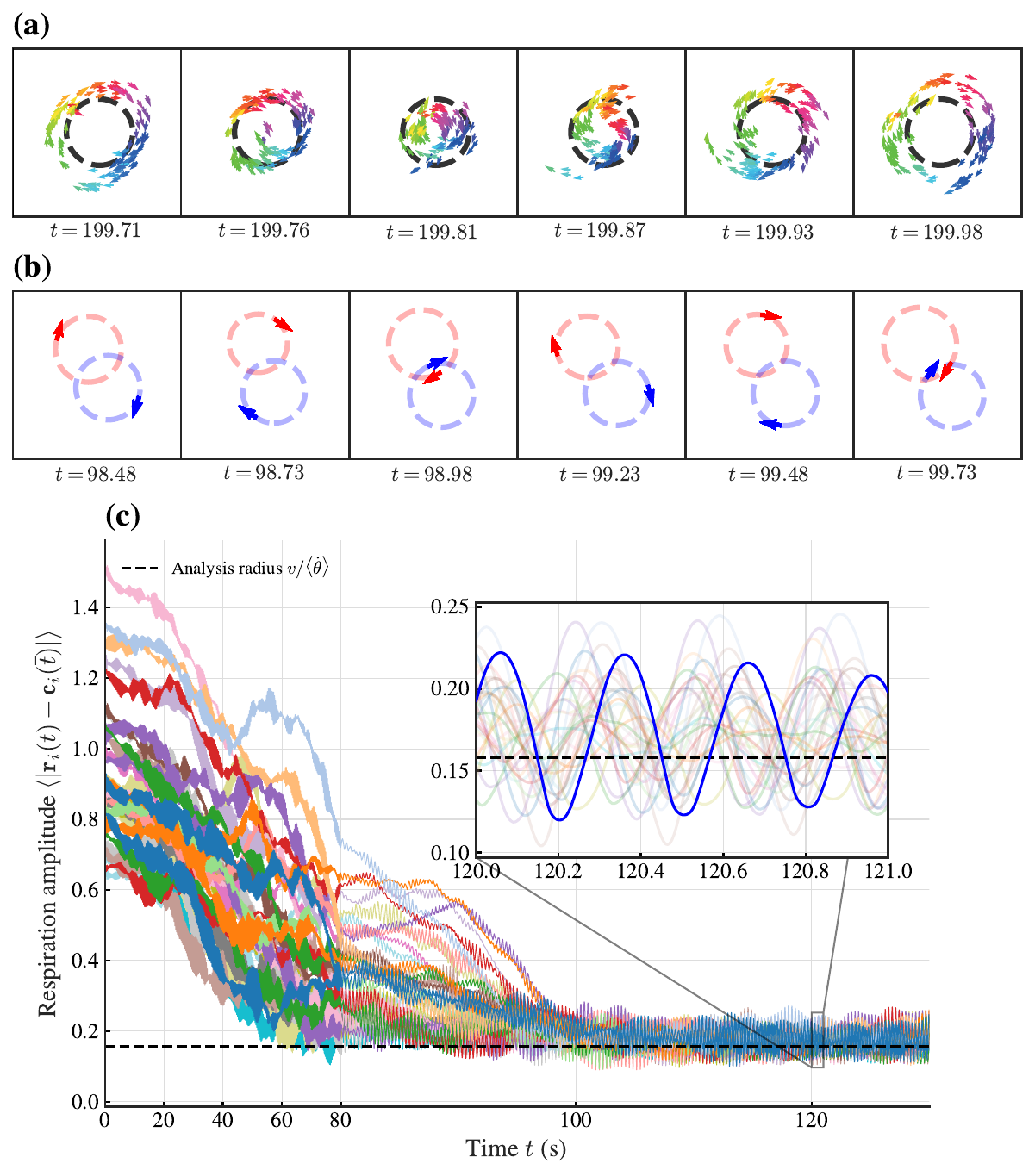}
        \caption{
            \label{fig:respiration}
            (a, b) Respiration dynamics of vortex (a, $\alpha=0.6\pi$) and double clustered (b, $\alpha=0.9\pi$) unit cell. 
            In (b), dashed circle indicates vortex cell volume. In (k), red/blue arrows show phase-divided clusters; dashed circles show estimated trajectories from mean instantaneous frequencies $\langle \dot{\theta}_i\rangle $.
            (c) Respiration amplitude over time for $\alpha=0.6\pi$.
            Other parameters are same as in Fig.~\ref{fig:snapshotsAndPhaseDiagram}.
        }
\end{figure}

{\it Respiration-like Motions.}
To better understand the spatiotemporal patterns in lattice states, we analyze the dynamics of individual unit cells in the vortex and dual-cluster lattices (Fig.~\ref{fig:snapshotsAndPhaseDiagram}(j, k)). 
In the vortex lattice (Fig.~\ref{fig:snapshotsAndPhaseDiagram}(j)), particles undergo a respiration-like motion, periodically expanding and contracting around an average radius of $v/\langle \dot{\theta} \rangle$, where $\langle \dot{\theta} \rangle$, where the mean frequency in incoherent states is
\begin{equation}
    \begin{aligned}
        \langle \dot{\theta}\rangle &=-K\sin \alpha +\frac{K}{2\pi}\int_0^{2\pi}{\mathrm{d}\theta ^{\prime}\sin \left( \theta ^{\prime}-\theta +\alpha \right)}\\
        &=-K\sin \alpha\;.\\
    \end{aligned}  
    \label{eq:averageFrequency}
\end{equation}
This oscillation is seen in the particle-to-cell-center distance $\left| \mathbf{r}_i\left( t \right) -\mathbf{c}_i\left( \bar{t} \right) \right|$, as illustrated in Fig.~\ref{fig:snapshotsAndPhaseDiagram}(l), with $\mathbf{c}_i\left( \bar{t} \right)$ the time-averaged cell center. 
The dual-cluster lattice (Fig.~\ref{fig:snapshotsAndPhaseDiagram}(k)) shows greater order: two condensed particle groups rotate in opposite directions around the unit cell center on near-fixed circular paths (dashed circles), with radii also set by their mean effective frequencies $\langle \dot{\theta}_i\rangle$, as indicated by the dashed circles in Fig.~\ref{fig:snapshotsAndPhaseDiagram}(k). These respiration modes arise from a Hopf–Turing bifurcation, as discussed in the section {\it Critical transition and mechanism}.

\begin{figure}
    \centering
    \includegraphics[width=7.5cm]{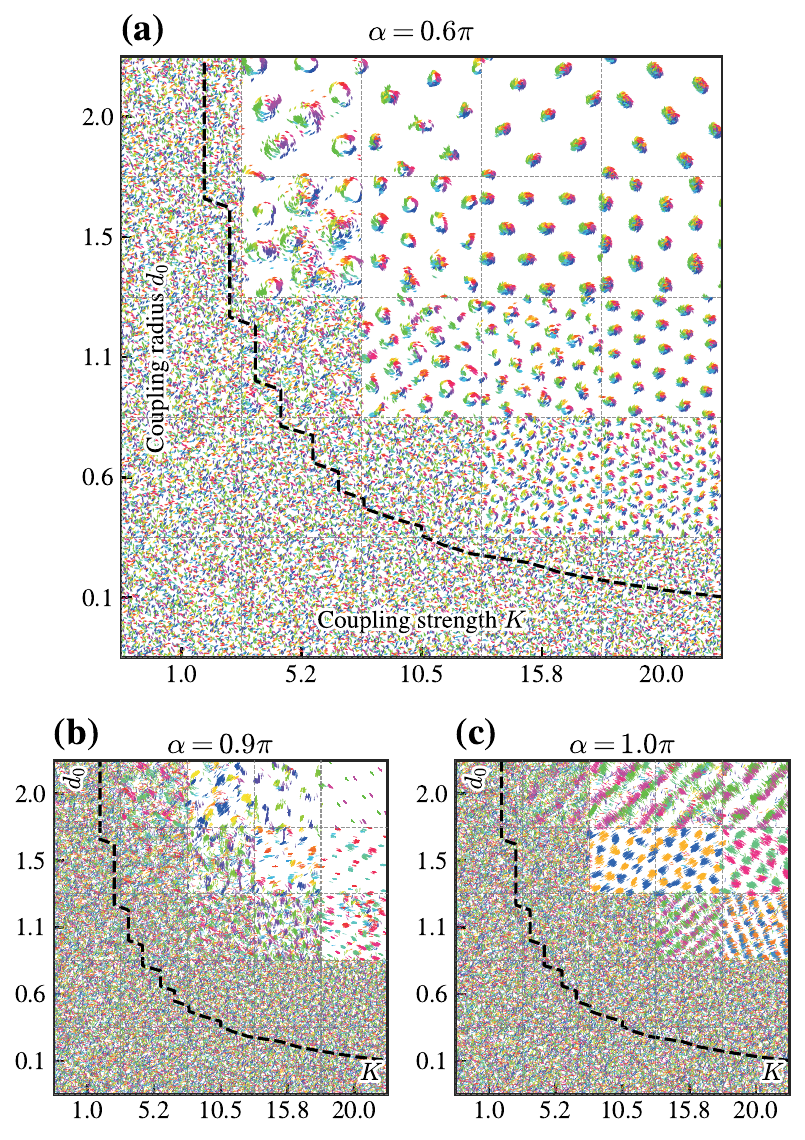}
    \caption{
        \label{fig:d0KphaseDiagram}
        Phase diagrams in the $(K, d_0)$ parameter space for different frustration $\alpha$ ($L=7, N=2000$, $v=3$). Black dashed lines mark boundaries (Eq. \eqref{eq:criticalLineOfKD0}) between dominant and recessive lattice states.
    }
\end{figure}

{\it Dominant-Recessive Lattice.}
As discussed above, the lattice state emerges when the frustration $\alpha$ exceeds the critical point $\pi/2$, with $\alpha$ primarily governing the topology of the lattice pattern.
However, the prominence of this lattice structure depends on the coupling strength $K$ and coupling radius $d_0$.
As illustrated in Fig.~\ref{fig:d0KphaseDiagram}(a)-(c), for a fixed $\alpha$, a region exists in the high $(K, d_0)$ parameter space where the lattice structure is dominant, characterized by large lattice constants (i.e., the average distance between neighboring unit cells) and large unit cell volumes (i.e., the area occupied by particles within a unit cell).
Within this region, the lattice constant increases with $d_0$ but remains largely unaffected by $K$. Conversely, the unit cell volume decreases as $K$ increases but shows minimal dependence on $d_0$.
Outside this region, the lattice structure becomes recessive, and the system tends toward a more uniform state. The boundary separating these two regimes can be approximated by \eqref{eq:criticalLineOfKD0}, which reflects a balance between the spatial periodicity of pattern formation and the spatial dynamics of the particles. The sawtooth shape of this boundary stems from the discrete nature of the wavenumber $k_n=2\pi n/L$ in a finite system with periodic boundary conditions.

{\it Critical transition and mechanism.}
In the thermodynamic limit $N\to \infty$, the state of the system in Eq.~\eqref{eq:totalDynamicsMeanField} can be characterized by the single-partial distribution $\rho \left( \mathbf{r},\theta ,t \right) $, which satisfies the continuity equation
\begin{equation}
    \frac{\partial \rho}{\partial t}=-v\mathbf{p}\left( \theta \right) \cdot \nabla \rho -\frac{\partial}{\partial \theta}\left\{ \mathcal{T} \left[ \rho \right] \rho \right\} \;,
    \label{eq:globalContinuityEquation}
\end{equation}
where $\mathcal{T}$ is the linear operator that has the form
\begin{equation}
    \begin{aligned}
        \mathcal{T} \left[ \rho \right] =&\frac{K}{A}\int_{L\times L}{\mathrm{d}^2\mathbf{r}^{\prime}\int_0^{2\pi}{\mathrm{d}\theta ^{\prime}\rho \left( \mathbf{r}^{\prime},\theta ^{\prime},t \right)}}\\
        &\times \Theta \left( d_0 - \left| \mathbf{r}^{\prime}-\mathbf{r} \right| \right) \left[ \sin \left( \theta ^{\prime}-\theta +\alpha \right) -\sin \alpha \right]\;,
    \end{aligned}
\end{equation}
where $\Theta(r) $ is Heaviside step function and  
\begin{equation}
    A=\int_{L\times L}{\mathrm{d}^2\mathbf{r}^{\prime}\int_0^{2\pi}{\mathrm{d}\theta ^{\prime}\rho \left( \mathbf{r}^{\prime},\theta ^{\prime},t \right) \Theta \left( d_0 - \left| \mathbf{r}^{\prime}-\mathbf{r} \right|\right)}}\;.
\end{equation}

One obvious solution of Eq.~\eqref{eq:globalContinuityEquation} is $\rho=(2\pi L^2)^{-1}$ representing a uniform disordered state. 
Inspired by the approaches in \cite{PhysRevE.109.024602}, the stability of such a solution can be investigated by considering a small perturbation,
\begin{eqnarray}
    \rho \left( \mathbf{r},\theta ,t \right) =\cfrac{1}{2\pi L^2}+\varepsilon \mathrm{e}^{\lambda \left( k \right) t+\mathrm{i}\mathbf{k}\cdot \mathbf{r}}\Phi \left( \theta \right) \;,
\end{eqnarray}
with $k=\left| \mathbf{k} \right|>0$, and linearizing the non-linear continuity equation \eqref{eq:globalContinuityEquation}, obtaining eigenvalues problem to compute the $\lambda(k)$ spectrum
\begin{equation}
    \left( \mathcal{L} _0-\mathrm{i}vk\mathcal{L} _1 \right) \Phi =\lambda \Phi \;,
\end{equation}
$\mathcal{L}_0$ is diagonal in the basis $\left\{ \mathrm{e}^{\mathrm{i}m\theta} \right\} _{m=-\infty}^{\infty}$,
\begin{equation}
    \mathcal{L} _0\Phi _m=\lambda _{m}^{\left[ 0 \right]}\mathrm{e}^{\mathrm{i}m\theta}\;,
\end{equation}
with the eigenvalues
\begin{equation}
    \lambda _{m}^{\left[ 0 \right]}\left( k \right) =\frac{KJ_1\left( kd_0 \right)}{kd_0}\left( \delta _{m,-1}\mathrm{e}^{\mathrm{i}\alpha}+\delta _{m,1}\mathrm{e}^{-\mathrm{i}\alpha} \right)  \;,
\end{equation}
where $J_1\left( x \right)$ is the Bessel function of the first kind. The operator $\mathcal{L}_1$ is defined as
\begin{equation}
    \mathcal{L} _1\mathrm{e}^{\mathrm{i}m\theta}=\frac{1}{2}\left( \mathrm{e}^{\mathrm{i}\left( m+1 \right) \theta -\mathrm{i}\vartheta}+\mathrm{e}^{\mathrm{i}\left( m-1 \right) \theta +\mathrm{i}\vartheta} \right)  \;,
\end{equation}
where $\vartheta$ is the forms $\mathbf{k}$ with the $x$ axis. Without the loss of generality, we can define the $x$ axis parallel to $\mathbf{k}$, and, therefore, take $\vartheta = 0$. 

For analytical convenience, truncating the expansion at order $M=2$ gives a 5th-degree characteristic polynomial $P_5(\lambda)$ with the following coefficients:
\begin{equation}
    \begin{array}{c}
	c_5=1,\;c_4=-\frac{2KJ_1\left( d_0k \right) \cos \alpha}{d_0k},\;c_3=\frac{K^2J_{1}^{2}\left( d_0k \right)}{d_{0}^{2}k^2}+k^2v^2\;,\\
	c_2=-\frac{Kkv^2J_1\left( d_0k \right) \cos \alpha}{d_0},\;c_1=\frac{3k^4v^4}{16},\;c_0=0\;.\\
\end{array}
\end{equation}
Since $\lambda =0$ is always a root, we analyze stability using the Routh–Hurwitz criterion on the reduced quartic polynomial $P_4\left( \lambda \right)$,  yielding the necessary and sufficient condition:
\begin{equation}
    \label{eq:instabilityCondition}
    J_1\left( d_0k \right) \cos \alpha <0\;.
\end{equation}
This shows stability depends only on $\alpha$ and $d_0$, not on $K$ and $v$.
Considering the range of $k$, $d_0$ only affect the unstable wave number without altering the qualitative results. And obviously, $\alpha=\pi/2$ is a critical value.

\begin{figure}
    \centering
    \includegraphics[width=7.5cm]{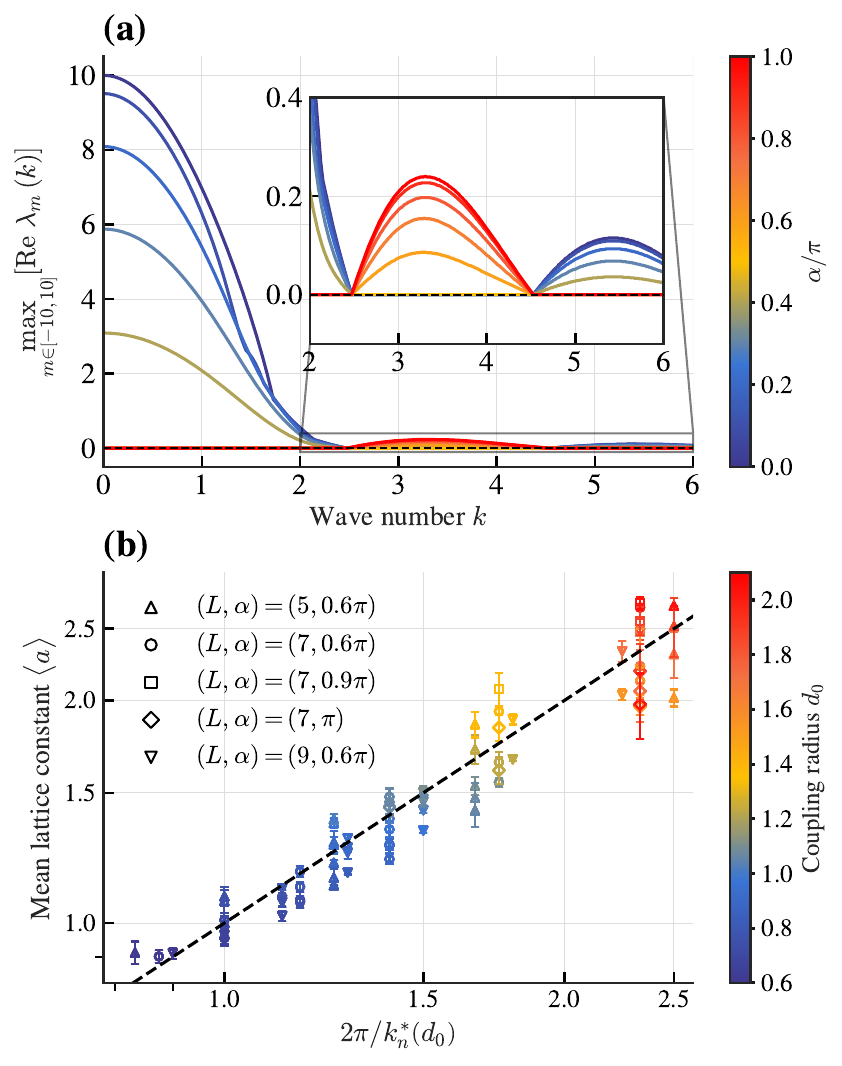}
    \caption{
        \label{fig:eigenvaluesAndLatticeConstants}
        (a) Computations of the $\lambda(k)$ with the largest real part as a function of the continuous wavenumber $k$ and $\alpha$ in the truncated basis $M=10$. Other parameters as in Fig.~\ref{fig:snapshotsAndPhaseDiagram}. 
        (b) Measured mean lattice constant $\langle a \rangle$ compared to theoretical prediction. Horizontal and vertical error bars represent standard deviations across $K$ and unit cells, respectively. The black dashed line indicates a 1:1 correspondence.
    }
\end{figure}

As computations of $\lambda(k)$ shown in Fig.~\ref{fig:eigenvaluesAndLatticeConstants}(a), for $\alpha < \pi/2$, instability occurs at k=0 (type III, no pattern formation, \cite{RevModPhys.65.851}).
For $\alpha > \pi/2$, analysis of $P_4^{\prime}(\lambda)$ reveals a II$_o$ instability (Hopf-Turing bifurcation), from complex roots with positive real parts, generating spatiotemporal patterns---here, lattice respiration motions. Thus, $\alpha = \pi/2$ is the critical point for lattice emergence.

Most notably, the most unstable wave number $k^{*}(d_0)$ is largely independent of $\alpha (>\pi/2)$, implying patterns near threshold have similar spatial periodicity. This matches the nearly constant $p_{\mathrm{std}}$ unchanged over the interval $\alpha \in (0.55\pi,0.9\pi)$ in Fig.~\ref{fig:snapshotsAndPhaseDiagram}(i) and the observed constant lattice spacing in simulations (Fig. ~\ref{fig:eigenvaluesAndLatticeConstants}(c)), predictable via wavelength $2\pi/k^{*}_n(d_0)$.

Significantly, apart from spatial periodicity, the volume of the unit cell is another important characteristic of the lattice state. 
For the vortex lattice near $\alpha = \pi/2$, phase is incoherent, and the mean effective rotational radius is approximately $v/|\dot{\theta}_i| \approx v/K$.
The lattice emergence condition balances pattern periodicity and particle dynamics, occurring when two unit cells tangent, i.e., cell diameter equals wavelength:
\begin{equation}
    \frac{2\pi}{k^*\left( d_0 \right)}=\frac{2v}{K}\;,
    \label{eq:criticalLineOfKD0}
\end{equation}
This matches the $(K, d_0)$ phase diagram (Fig.~\ref{fig:d0KphaseDiagram}(a)-(c)). 
When the wavelength is too short to accommodate the unit cell volume, the cells overlap, leading to macroscopic uniformity. 

{\it Conclusions.}
Frustrated alignment in the Vicsek-Kuramoto system can induce a resting hexagonal lattice structure in self-propelled particles, demonstrating that purely orientational interactions are sufficient for symmetric pattern formation without explicit spatial forces. This system exhibits a rich array of collective states: For low frustration parameters ($\alpha<\pi/2$), it achieves synchronized states such as global synchronization and swarming, whereas beyond the critical point ($\alpha > \pi/2$), it transitions into lattice patterns including vortex lattices and dual-cluster lattices, as visualized in the spatial configurations (Fig.~\ref{fig:snapshotsAndPhaseDiagram}). The critical transition at $\alpha = \pi/2$ is characterized by a Hopf-Turing bifurcation, leading to spatiotemporal patterns with respiration-like dynamics within the unit cells (Fig.~\ref{fig:respiration}). The dominance of the lattice structure depends on the coupling strength and radius, with a clear boundary in the $(K, d_0)$ parameter space (Fig.~\ref{fig:d0KphaseDiagram}), explained by balancing pattern periodicity and particle dynamics (Eq.~\eqref{eq:criticalLineOfKD0}). This work establishes frustration as a design principle for tuning lattice topology in active matter systems, but leaves open questions regarding the stability of pattern selection and the prediction of respiration frequencies, warranting further investigation to characterize the nature of these emergent structures.
    
{\it Acknowledgments.}
We are grateful to Professor Yunyun Li at Tongji University for fruitful discussions.
This work is partially supported by National Natural Science Foundation of China (Nos. 12375031 and 11875135).

\bibliography{ref}

\end{document}


\setlength{\parindent}{0.5cm}

\title{Supplementary information for \\ \enquote{Swarming Lattice in the Frustrated Vicsek-Kuramoto Model}}

\maketitle

\section{Nonidentical particles}

\begin{figure}[H]
        \centering
        \includegraphics[width=0.95\textwidth]{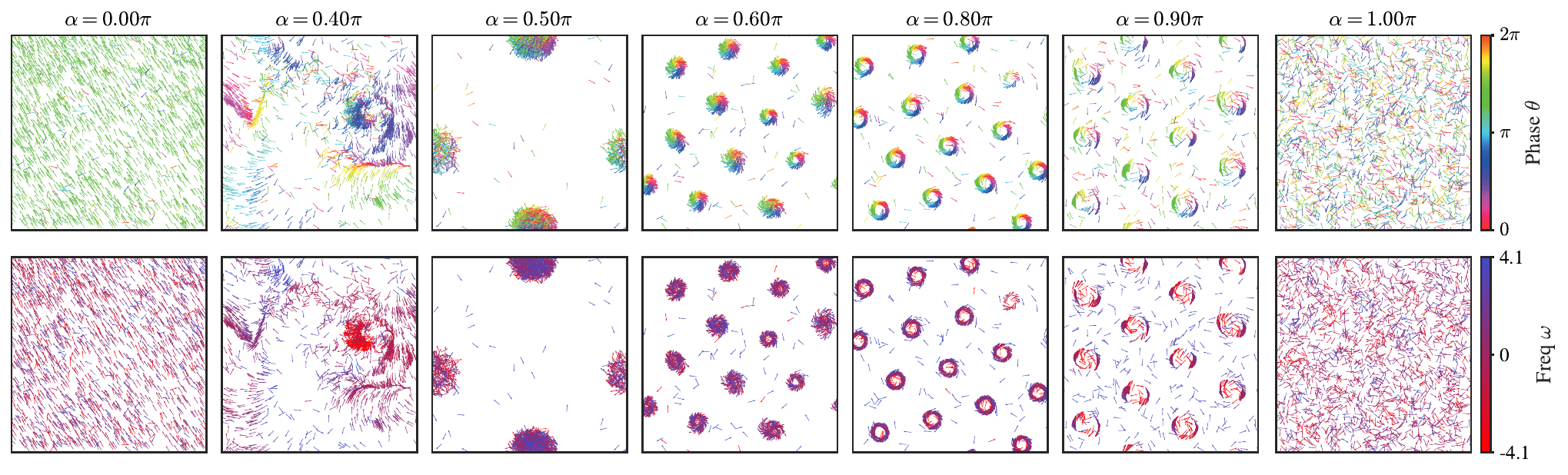}
        \caption{
            \label{fig:cauchyDistSnapshots}
            \justifying
            Snapshots of the system with nonidentical particles drawn from a Lorentzian distribution with scale parameter $\gamma = 1$ at different frustration $\alpha$. 
            Top: local orientation color map.
            Bottom: local frequency color map.
            Other parameters are same as in Fig.~1 of the main text.
        }
\end{figure}

\begin{figure}[H]
        \centering
        \includegraphics[width=0.95\textwidth]{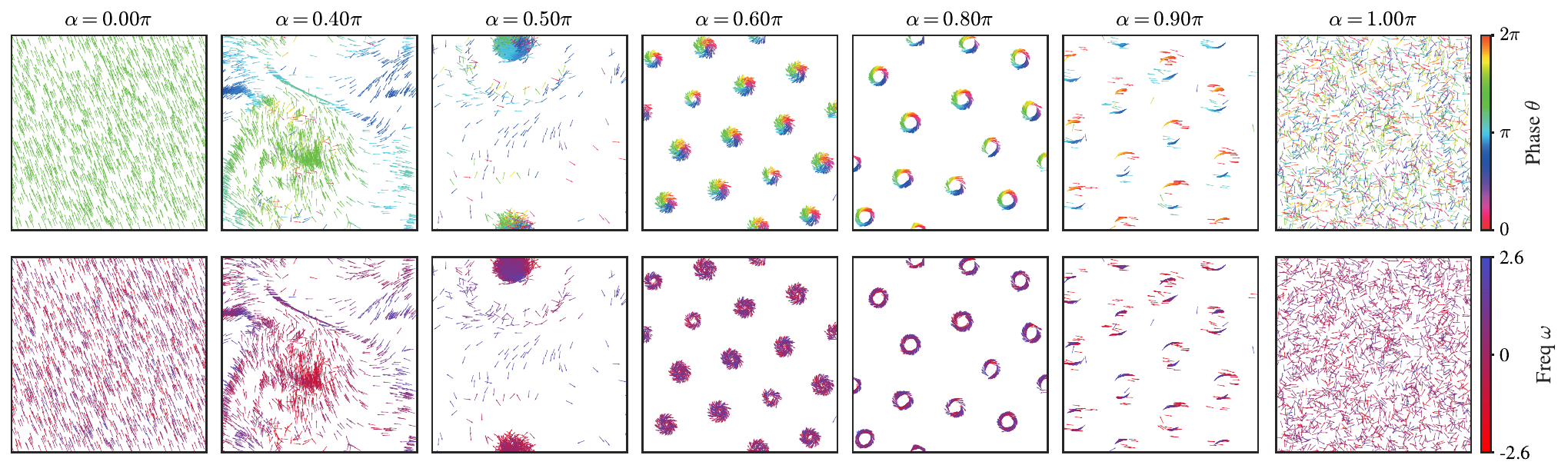}
        \caption{
            \label{fig:gaussianDistSnapshots}
            \justifying
            Snapshots of the system with nonidentical particles drawn from a Gaussian distribution with mean 0 and standard deviation $\sigma = 1$ at different frustration $\alpha$. 
            Other parameters are same as in Fig.~1 of the main text.
        }
\end{figure}

\begin{figure}[H]
        \centering
        \includegraphics[width=0.95\textwidth]{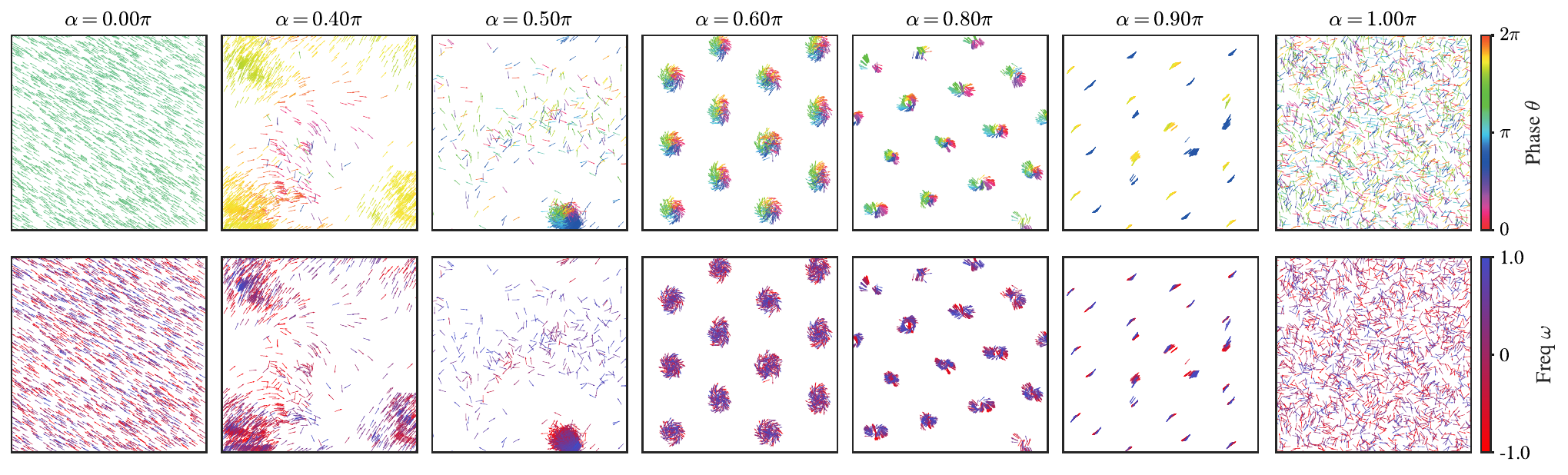}
        \caption{
            \label{fig:uniformDistSnapshots}
            \justifying
            Snapshots of the system with nonidentical particles drawn from a uniform distribution with mean 0 and half-width $a = 1$ at different frustration $\alpha$. 
            Other parameters are same as in Fig.~1 of the main text.
        }
\end{figure}

In main text, we have focused on the case of identical particles. Here, we extend our analysis to nonidentical particles by introducing intrinsic frequencies $\omega_i$ for each particle $i$. The dynamics of the system is then governed by the following set of equations:
\begin{subequations} 
    \label{eq:totalDynamicsMeanField}
    \begin{align}
        \dot{\mathbf{r}}_i&=v\mathbf{p}\left( \theta _i \right)\;\label{eq:dotR},
        \\
        \dot{\theta}_i&=\omega _i+\frac{K}{\left| A_i \right|}\sum_{j\in A_i}{\left[ \sin \left( \theta _j-\theta _i+\alpha \right) -\sin \alpha \right]}\;\label{eq:dotTheta},
    \end{align}
\end{subequations}
where $\omega_i$ is the intrinsic frequency of particle $i$, drawn from a distribution $g(\omega)$ with zero mean and finite variance. Firstly, we consider a Lorentzian distribution $g(\omega)=\frac{\gamma}{\pi}\frac{1}{\omega^2+\gamma^2}$ with scale parameter $\gamma$. 
As shown in Fig.~\ref{fig:cauchyDistSnapshots}, the lattice structure is robust against the introduction of intrinsic frequencies, and the lattice structure still emerges as frustration $\alpha$ increases. We have also tested Gaussian distribution $g(\omega)=\frac{1}{\sigma\sqrt{2\pi}}\exp\left(-\frac{\omega^2}{2\sigma^2}\right)$ with standard deviation $\sigma$ (Fig.~\ref{fig:gaussianDistSnapshots}) and uniform distribution $g(\omega)=\frac{1}{2a}$ for $\omega\in[-a,a]$ with half-width $a$ (Fig.~\ref{fig:uniformDistSnapshots}), and observed similar lattice formation as frustration increases. This suggests that the formation of lattice is a robust phenomenon that persists even in the presence of nonidentical particles with intrinsic frequencies.

\section{Noise effects}
\begin{figure}[H]
    \centering
    \subcaptionbox{
        $\alpha=0.6\pi$
    }[0.32\linewidth]{
        \includegraphics[width=\linewidth]{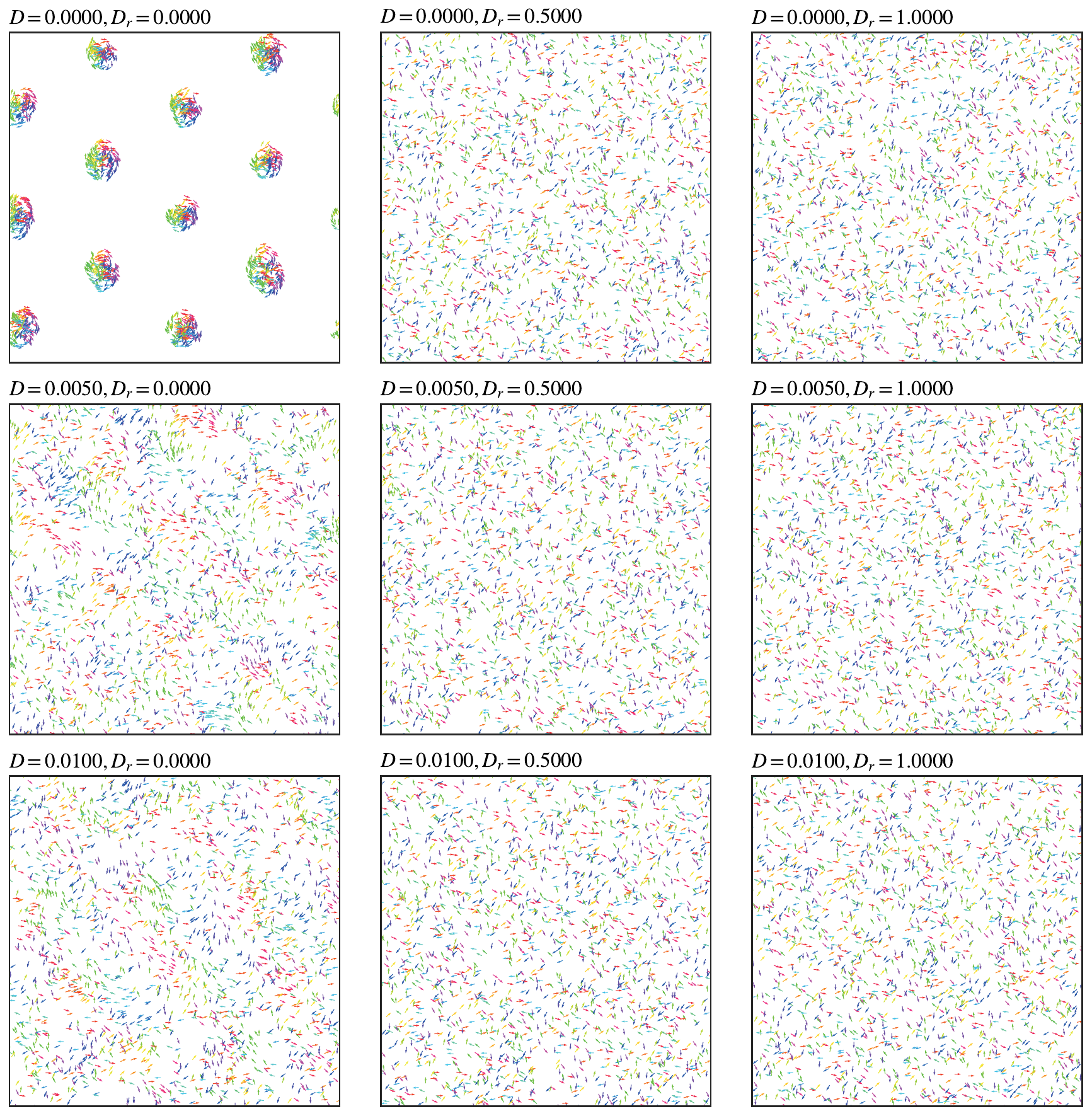}
    }
    \hfill
    \subcaptionbox{
        $\alpha=0.6\pi$
    }[0.32\linewidth]{
        \includegraphics[width=\linewidth]{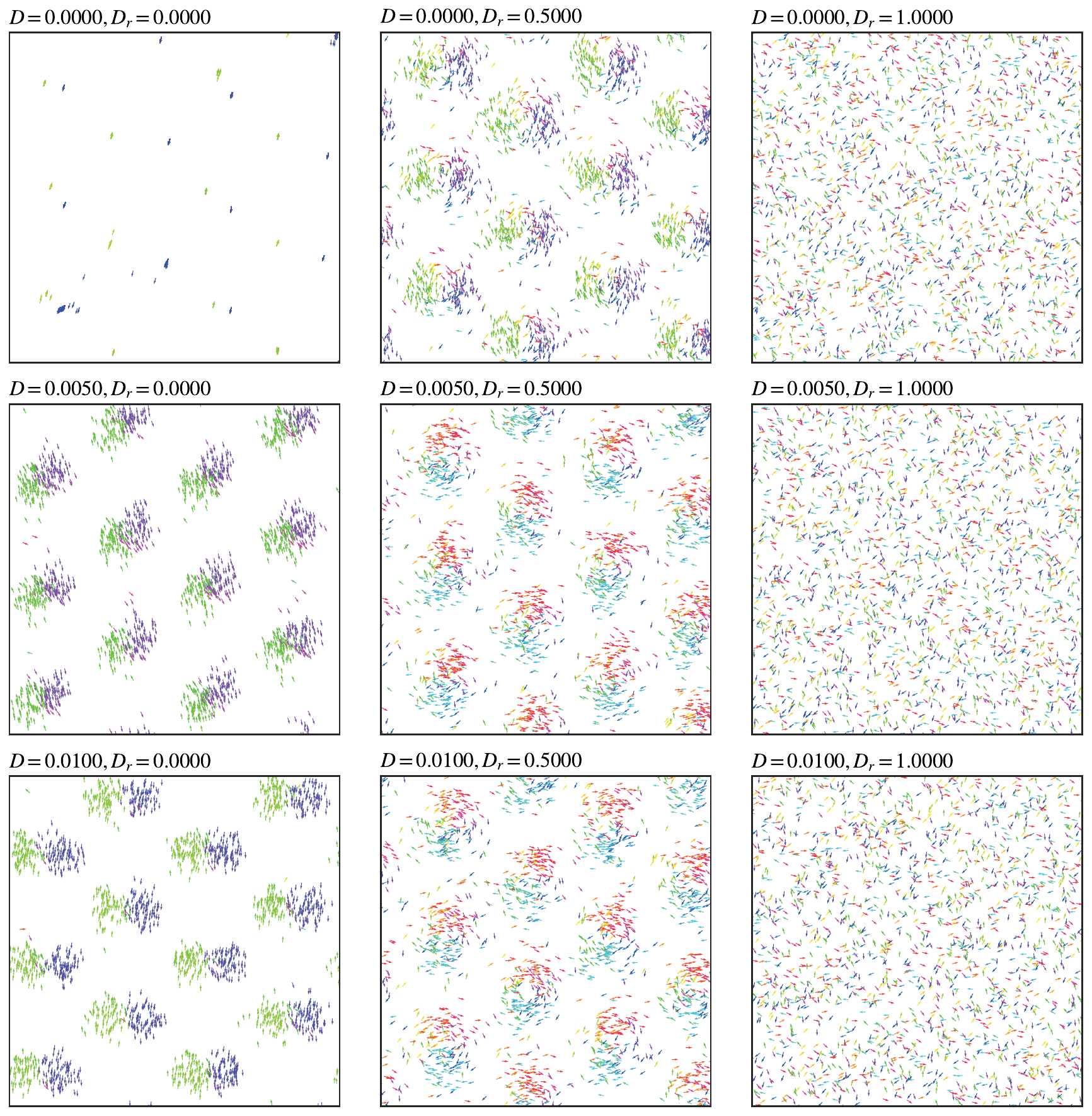}
    }
    \hfill
    \subcaptionbox{
        $\alpha=\pi$
    }[0.32\linewidth]{
        \includegraphics[width=\linewidth]{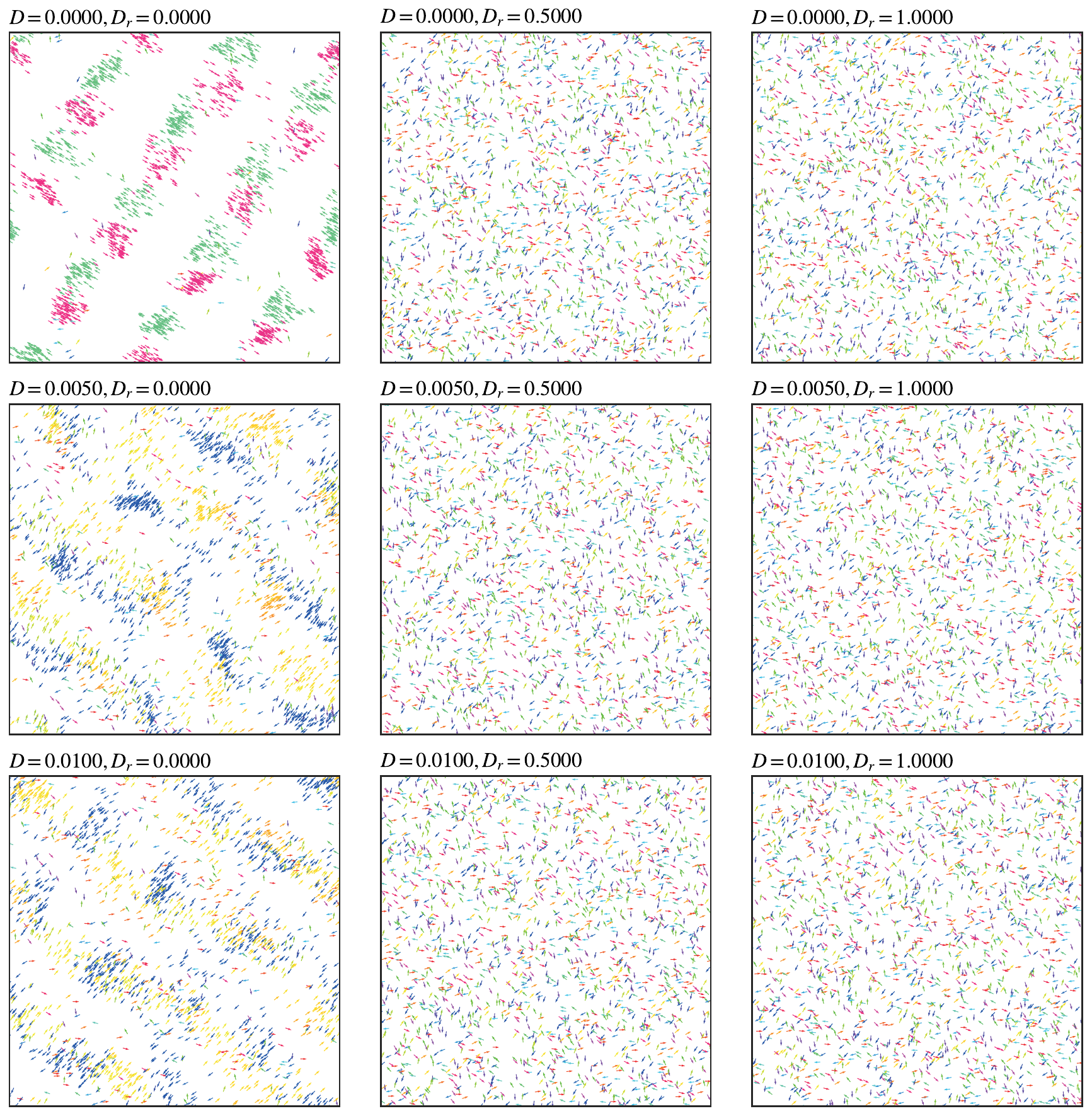}
    }
    \caption{
        \label{fig:noiseEffects}
        \justifying
        Snapshots of the system with varying levels of translational noise $D$ and rotational noise $D_r$ at different frustration $\alpha$. 
        Other parameters are same as in Fig.~1 of the main text.
    }
\end{figure}

To investigate the effects of noise on the formation of lattice structures, we introduce Gaussian white noise terms into the dynamics of both position and orientation of the particles. The modified equations of motion are given by:
\begin{subequations} 
    \label{eq:totalDynamicsWithNoise}
    \begin{align}
        \dot{\mathbf{r}}_i&=v\mathbf{p}\left( \theta _i \right)+\sqrt{2D}\boldsymbol{\xi}_i(t)\;\label{eq:dotRWithNoise},
        \\
        \dot{\theta}_i&=\frac{K}{\left| A_i \right|}\sum_{j\in A_i}{\left[ \sin \left( \theta _j-\theta _i+\alpha \right) -\sin \alpha \right]}+\sqrt{2D_r}\eta_i(t)\;\label{eq:dotThetaWithNoise},
    \end{align}
\end{subequations}
where $\boldsymbol{\xi}_i(t)$ and $\eta_i(t)$ are independent Gaussian white noise terms with zero mean and unit variance, representing translational and rotational noise, respectively. The parameters $D$ and $D_r$ denote the translational and rotational diffusion coefficients. As shown in Fig.~\ref{fig:noiseEffects}, the lattice structures remain robust against moderate levels of both translational and rotational noise. However, as the noise levels increase significantly, the lattice structures begin to deteriorate, leading to disordered states. This indicates that while the formation of lattice structures is resilient to noise, there exists a threshold beyond which the ordered patterns are disrupted.

